\documentclass[10pt]{iopart}

\usepackage{graphicx}  
\usepackage{dcolumn}
\usepackage{bm}

\begin{document}

\title[Search for  half-metallic ferrimagnetism  in V-based
Heusler alloys]{Search for  half-metallic ferrimagnetism  in
V-based  Heusler alloys Mn$_2$VZ (Z$=$Al, Ga, In, Si, Ge, Sn)}

\author{K \"Ozdo\~gan\dag, I Galanakis\ddag, E \c Sa\c s\i
o\~glu\S, and B Akta\c s\dag}

\address{\dag\ Department of Physics, Gebze Institute of Technology,
Gebze, 41400, Kocaeli, Turkey }

\address{\ddag\ Department of Materials Science, School of Natural
  Sciences, University of Patras,  GR-26504 Patra, Greece}

\address{\S\ Max-Planck-Institut f\"ur Mikrostrukturphysik,
D-06120 Halle, Germany}

\ead{kozdogan@gyte.edu.tr,i.galanakis@fz-juelich.de,ersoy@mpi-halle.de}

\begin{abstract}

Using a state-of-the-art full-potential electronic structure
method within the local spin density approximation, we study the
electronic and magnetic structure of Mn$_2$V-based full Heusler
alloys: Mn$_2$VZ (Z=Al, Ga, In, Si, Ge, and Sn). We show that
small expansion of the calculated theoretical equilibrium lattice
constants restores the half-metallic ferrimagnetism in these
compounds. Moreover a small degree of disorder between the V and Z
atoms, although iduces some states within the gap, it preserves
the Slater-Pauling behaviour of the spin magnetic moments and the
alloys keep a high degree of spin-polarisation at the Fermi level
opening the way for a half-metallic compensated ferrimagnet.
\end{abstract}


\section{Introduction}

Recently rapid development of  magneto electronics intensified the
researches on the ferromagnetic materials that are suitable for
the spin injection into a semiconductor\cite{ohno}. One of the
promising classes of materials are the half--metallic
ferrimagnets, i.e., compounds for which  only one spin channel
presents a gap at the Fermi level, while the other has a metallic
character, leading to 100\% carrier spin-polarization at $E_F$
\cite{Zutic,deBoeck}. The half--metallic ferromagnetism (HMF) was
discovered by de Groot \textit{et al.} in 1983 when studying the
band structure of semi Heusler compound NiMnSb \cite{deGroot}.
Ishida \textit{et al.} have proposed that also the full-Heusler
alloys compounds of the type Co$_2$MnZ, where Z stands for Si and
Ge, are half-metals \cite{Ishida}. Since then many others have
been predicted on the basis of ground state calculations, such as
magnetic oxides (CrO$_2$ and Fe$_3$O$_4$), colossal
magnetoresistance materials (Sr$_2$FeMoO$_6$ and
La$_{0.7}$Sr$_{0.3}$MnO$_3$) \cite{Soulen}. Finally the diluted
magnetic semiconductors (Ga$_{1-x}$Mn$_x$As) and  MnAs and CrAs in
the  zincblende crystal structure  have attracted a lot of
attention \cite{Freeman,Akai,Akinaga}.

Heusler alloys have been particularly interesting  systems because
they exhibit much higher ferromagnetic Curie temperature than
other half--metallic materials \cite{Webster}. Among the other
properties useful for the applications are  the crystal structure
and lattice matching compatible with zinc-blende semiconductors
used industrially  \cite{lattice_match_1,lattice_match_2}.

Mn$_2$VAl received much experimental and theoretical attention.
The neutron diffraction experiment by Itoh et al. \cite{itoh}
revealed the  ferrimagnetic ordering in this  compound with a Mn
magnetic moment of $1.5\pm 0.3 \mu_B$ and a V moment of $-0.9
\mu_B$. Yoshida \textit {et al.} studied the  magnetic  and
structural properties  of Mn$_2$V$_{1+x}$Al$_{1-x}$ as a function
of $x$ \cite{Yoshida}. The authors  found that the structure
preserved the   L2$_1$-type phase for $-0.3<x<0.2$ with linearly
varying saturation moment. For $x\geq 0.2$  a structural
transformation from L2$_1$-phase to a disordered B2-type phase has
been detected. Recently Jiang {\textit et al.} examined the
magnetic structure of Mn$_2$VAl by X-ray diffraction and
magnetization measurements \cite{jiang}. They found that Mn$_2$VAl
was nearly half-metallic with the total magnetic moment of
1.94$\mu_B$ at 5 K. The loss of half-metallic character was
attributed to the small amount of disorder in  V and Al
sublattices.

The electronic  structure of Mn$_2$VAl  has  been studied for the
first  time  by Ishida \textit {et al.} \cite{Ishida2}. The
authors used local-density approximation (LDA) to the density
functional theory and showed that ground state of Mn$_2$VAl  was
close to half-metallicity. Recently a detailed theoretical study
of the magnetism of Mn$_2$VAl was reported by Weht and  Pickett
\cite{ruben} who used the generalized gradient approximation (GGA)
for the exchange correlation potential and have shown that
Mn$_2$VAl is a half-metallic ferrimagnet with the atomic moments
of 1.5$\mu_B$ and -0.9$\mu_B$ for Mn and V atoms respectively, in
very good agreement with experiment. The Fermi level was found to
lie in the minority spin band. In 2005 \c Sa\c s\i o\~glu {\textit
et al.} studied the exchange interaction and Curie temperature in
the Mn$_2$VZ (Z=Al,Ge) half--metallic compounds and showed that
the antiferrimagnetic coupling between the V  and  Mn atoms
stabilizes the ferromagnetic alignment of the Mn spin moments
\cite{ersoy}.

\section{Motivation}

Half-metallic ferrimagnetic  materials, like FeMnSb or the
Mn$_2$VZ compounds, are much more desirable than  their
ferromagnetic counterparts in magnetoelectronics applications.
This is mostly due to the fact that the small value of the total
magnetic moment in these systems provides additional advantageous.
For example they do not  give rise to strong stray fields in
devices or  are less affected by the  external magnetic fields.

The ideal case for applications would be a half-metallic
antiferromagnet like CrMnSb. It is a special antiferromagnet in
the sense that the majority spin and minority spin densities of
states are not identical, as for common antiferromagnets and the
material is better described as a fully compensated ferrimagnet,
having a magnetic moment that is, due to the half-metallic
character, precisely equal to zero. Such a half-metallic
antiferromagnet would be a very interesting magnetoelectronic
material since it would be a perfectly stable spin-polarized
electrode in a junction device. And moreover if used as a tip in a
spin-polarized STM, it would not give rise to stray flux, and
hence would not distort the domain structure of the soft-magnetic
systems to be studied. Unfortunately, CrMnSb does not crystallize
in the ordered C1$_b$ crystal structure adopted by the
half-Heusler alloys. However, these results show that such an
important magnetoelectronic material could exist. Van Leuken and
de Groot have recently suggested a possible route towards a
half-metallic antiferromagnet starting from the semiconducting
C1$_b$-type compound FeVSb \cite{Leuken}. It is isoelectronic with
the non-existing half-metallic antiferromagnet CrMnSb. A 12.5 \%
substitution of Mn for V, and (in order to keep the system
isoelectronic, In for Sb) was predicted to already yield
half-metallic ferrimagnetism, with local Mn moments of about 2.3
$\mu_B$  and a band gap of about 0.35 eV.

In the  present paper,  we systematically study  the electronic
and magnetic structure of V--based full Heusler alloys Mn$_2$VZ
(Z=Al, Ga, In, Si, Ge, and Sn) to search for new half-metallic
ferrimagnetic candidates and search for way to decrease their
spin-moment in order to reach the ideal case of the half-metallic
compensated ferrimagnet. Among the systems studied Mn$_2$VZ (Z=Al,
Ga, In and Sn) are predicted to be nearly half-metallic at
theoretical equilibrium lattice constants. We demonstrate that
half-metallicity can be achieved in Mn$_2$VZ (Z=Al, Se, Ge, Sn) by
expanding the lattice parameter by a few percent  and  can be
maintained for  a small range  of lattice constants. Finally the
effect of disorder, and more specifically the intermixing of V and
$sp$ atoms, on the half-metallicity and spin polarization of the
Mn$_2$VAl is discussed.

The  paper is  organized as follows. In section \ref{sec3} we
present the calculational approach. Section \ref{sec4} contains
the results and the discussion on the total energy calculations
and section \ref{sec5} is devoted to effect of disorder. Finally
in section \ref{sec6} we summarise and conclude.

\section{Calculational method} \label{sec3}

\begin{figure}
\begin{center}
\includegraphics[scale=0.34]{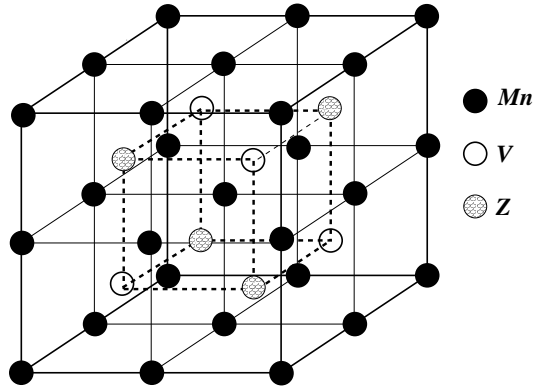}
\end{center}
\caption{Schematic representation of the $L2_1$ structure adapted
by the full Heusler compounds. The lattice consists of four
interpenetrating fcc sublattices with the positions $(0,0,0)$ and
$(\frac{1}{2},\frac{1}{2},\frac{1}{2})$  for the Mn and
$(\frac{1}{4} \frac{1}{4}\frac{1}{4})$ and
$(\frac{3}{4},\frac{3}{4},\frac{3}{4})$ for the V and \textit{sp}
(Z) atoms, respectively.} \label{fig_lattice}
\end{figure}

Full Heusler compounds crystallize in the $L2_1$--type structure
(see figure~\ref{fig_lattice}). The lattice consists of four
interpenetrating fcc sublattices and is characterized by the
formula X$_2$YZ. In the compounds under study,  Mn atoms occupy
the X sites in contrast to the majority of the Heusler alloys for
which Mn atom usually enters as the Y element.  Vanadium atoms
occupy the Y sites while the Z site corresponds to the $sp$
element (in our case Al, Ga, In, Si, Ge or Sn).

The electronic structure calculations are  performed using the
full--potential nonorthogonal local--orbital minimum--basis band
structure scheme (FPLO) \cite{koepernik}. We use the scalar
relativistic formulation and thus the spin-orbit coupling is not
taken into account. The exchange--correlation potential is chosen
in the local spin density approximation (LSDA) \cite{perdew}. The
self-consistent potentials were calculated on a
$20\times20\times20$ \textbf{k}-mesh in the Brillouin zone, which
correspond to 256 k points in the irreducible Brillouin zone. The
set of valance orbitals in the FPLO calculations were selected as
3s, 3p, 4s, 4p, 3d for Mn, Ga, Ge and V, 4s, 4p, 5s, 5p, 4d for In
and Sn and 2s, 2p, 3s, 3p, 3d for Al and Si . All lower states
were treated as core states.

Disorder in  Mn$_2$V$_{1-x}$Al$_{1+x}$ was  introduced  within the
coherent potential approximation (CPA) implemented in FPLO code
\cite{koepernik}.

\section{Half--metallic ferrimagnetism in Mn$_2$VZ (Z=Al,Ga, In, Si, Ge, Sn) }
\label{sec4}

\subsection{Total energy and lattice  parameter}

\begin{figure}
\begin{center}
\includegraphics[scale=0.5]{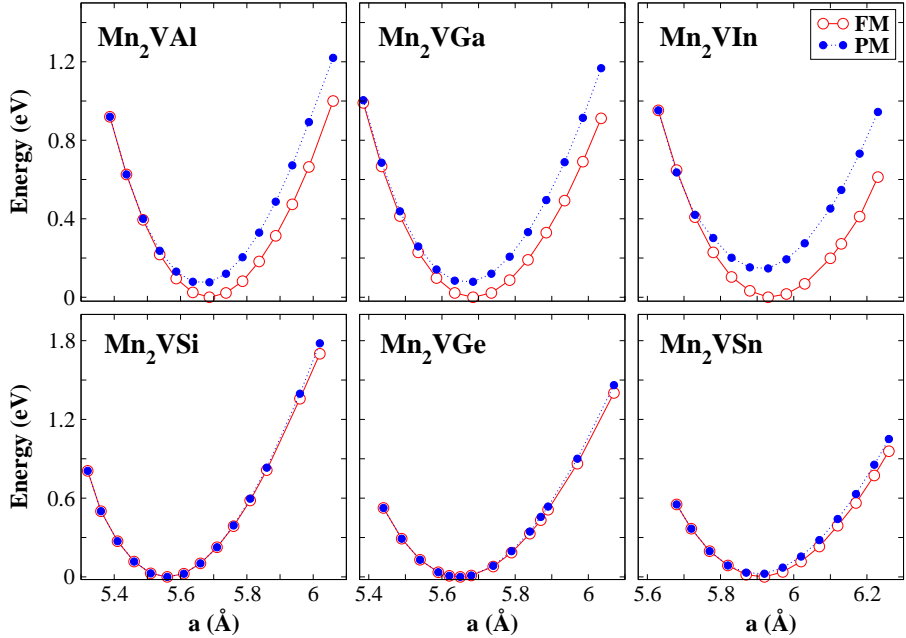}
\end{center}
\caption{Calculated total energy for all studied compounds as a
function of lattice constant for both the non-magnetic (PM) and
ferrimagnetic (FM) cases. The total energies have been rescaled so
that the zero energy corresponds to the calculated equilibrium
lattice constant. Spheres represent calculated values.}
\label{fig2}
\end{figure}

The first task, which we undertook, was to calculate how the total
energy changes with the lattice constant and thus to find the
theoretical equilibrium lattice parameter. We performed
calculations both for the non-magnetic and the ferrimagnetic cases
and we summarize our results in figure \ref{fig2}. There is a
clear difference when the $sp$ atom belong to the IIIB column of
the periodic table (Al, Ga or In) having three valence electrons
with respect to the case of a IVB column $sp$ element (Si, Ge or
Sn) with four valence electrons. For Mn$_2$VAl(Ga or In)it is
clear that the ferrimagnetic state is the most stable one and the
stability (relative energy difference between the FM and PM
curves) increases as the $sp$ element is heavier reaching a value
of around 0.15 eV for Mn$_2$VIn at the equilibrium lattice
constant. In table~\ref{table1} we  have gathered the equilibrium
lattice constants, the difference in energy between the FM and PM
states and the atom-resolved and total spin moments. Interestingly
Mn$_2$VAl has a similar lattice constant to Mn$_2$VGa  and one has
to go to Mn$_2$VIn to see an important difference in the
calculated equilibrium lattice constant. Note also that the total
spin moments are not integers and thus the system is not
half-metallic at the equilibrium lattice constants.

\begin{table}
\caption{The calculated equilibrium lattice parameters (in
ferrimagnetic state), the energy difference $\Delta E$ between the
ferrimagnetic and  non-magnetic state at the equilibrium lattice
constant, and the atom-resolved spin magnetic moments in $\mu_B$
for the Mn$_2$VZ (Z=Al, Ga, In, Si, Ge, Sn) compounds. Last column
is the total spin magnetic moment in the primitive cell. Note that
for Si and Ge compounds we could not converge a ferrimagnetic
calculation at the equilibrium lattice constant (see text).)}
\begin{indented}
 \item[]
 \begin{tabular}{lcccccc}
  \hline \hline
Compound &  $a(\AA)$ &   $\Delta E(eV)$  &  $m_\mathrm{Mn}$ &
$m_\mathrm{V}$ &
$m_\mathrm{Z}$ & $m_\mathrm{Cell}$  \\
  \hline
Mn$_2$VAl & 5.687     &  0.076  & -1.264 &  0.581 &  0.017 & -1.93  \\
Mn$_2$VGa & 5.685     &  0.079  & -1.254 &  0.603 &  0.040 & -1.865  \\
Mn$_2$VIn & 5.930     &  0.140  & -1.404 &  0.853 &  0.038 & -1.917  \\
Mn$_2$VSi & 5.560     &  0.000  & 0.000 &  0.000 &  0.000 & 0.000  \\
Mn$_2$VGe &  5.650    &  0.000  & 0.000 &  0.000 &  0.000 & 0.000  \\
Mn$_2$VSn & 5.920     &  0.025  & -0.696 &  0.447 &  0.024 & -0.921  \\
\hline \hline
\end{tabular}
\end{indented}
\label{table1}
\end{table}

The case of Mn$_2$VSi(Ge or Sn) compounds is more interesting.
Here the non-magnetic and ferrimagnetic configurations are almost
degenarated. As we can see in figure~\ref{fig2} around the
equilibrium lattice constants it is impossible to distinguish the
black spheres representing the non-magnetic state form the empty
ones representing the ferrimagnetic state. Exactly at the
equilibrium lattice constant we were not able to converge the
ferrimagnetic solution for Mn$_2$VSi and Mn$_2$VGe compounds and
thus in table \ref{table1} we give only the equilibrium lattice
constant. On the other hand in the case of the Mn$_2$VSn compound
the ferrimagnetic state is energetically preferable with respect
to the non-magnetic one but the energy difference at the
equilibrium lattice constant is only 0.025 eV much smaller than
the Al-, Ga- and In-based compounds. Unfortunately only the
Mn$_2$VAl compound has been grown experimentally. But we expect
the magnetic properties of the Mn$_2$VSi(Ge or Sn) compounds to
strongly depend on the substrate and the growth conditions and the
latter compounds will very easily oscillate between the magnetic
and non-magnetic configurations.

Finally we should also note for the Mn$_2$VSi and Mn$_2$VGe
compounds that a very small expansion of the lattice parameter
with respect to the theoretical equlibrium lattice constant of the
order of 1\% stabilizes now the ferrimagnetic state leading to a
jump in the spin magnetic moments being the sign of a metamagnetic
transition. The intermediate region reguires a systemetic study
with fixed spin moment approach which is out of the scope of the
present work. For Mn$_2$VSi compound the Mn spin moment reaches a
value of -0.87 $\mu_B$ and V a moment of 0.35 $\mu_B$ with the
total spin moment being  -0.62 $\mu_B$. For Mn$_2$VGe the
corresponding moment are -0.82, 0.32 and -0.57 $\mu_B$. Similar
behaviour to Mn$_2$VSi and Mn$_2$VGe has been also found in the
Fe$_2$FeSi and Fe$_2$FeAl Heusler compounds \cite{Harmon} and in
Ni$_2$MnSn \cite{ersoy2}.

\subsection{Effect  of the lattice  parameter: appearance  of
half--metallicity}

All compounds under study are not half-metallic at their
theoretical equilibrium lattice constants. This is obvious from
table~\ref{table1} where the total spin moments in a cell are not
integer numbers. Galanakis and collaobrators have shown that in
the case of half-metallic full-Heusler alloys the total spin
moment in $\mu_B$ follows the relation $Z_t-24$, where $Z_t$ is
the total number of valence electrons, even for compounds with
less than 24 electrons as the ones under study
\cite{GalanakisFull}. Mn$_2$VAl(Ga or In) have 22 valence
electrons per cell and thus they should have a total spin moment
of -2 $\mu_B$ while the other three compounds Mn$_2$VSi(Ge or Sn)
have in total 23 valence electrons and a total spin moment per
cell of -1 $\mu_B$ in the case of half-metallicity. This is
obviously not the case in table \ref{table1}.

\begin{figure}
\begin{center}
\includegraphics[scale=0.5]{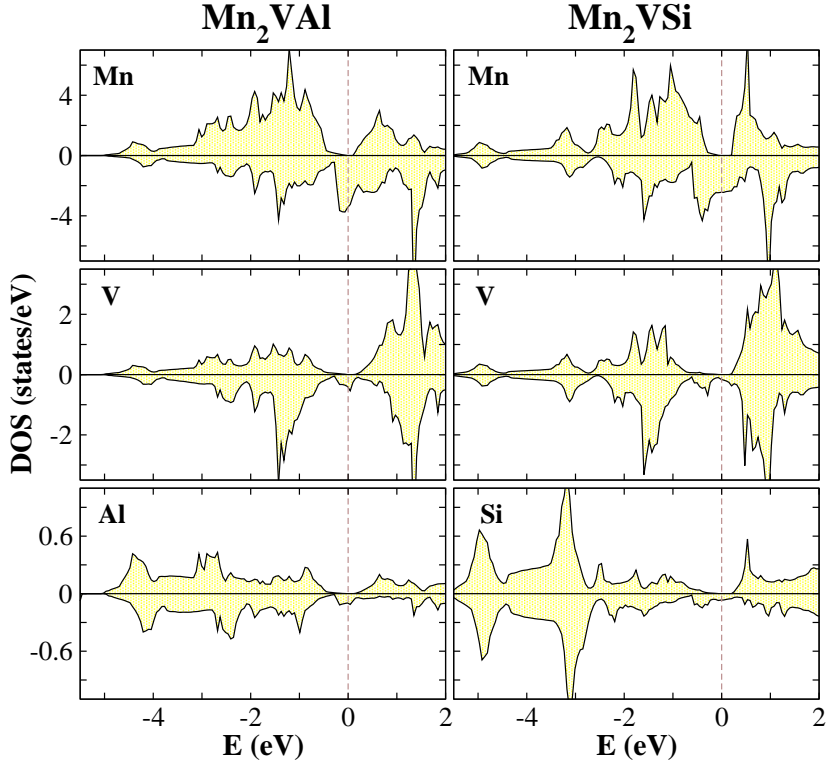}
\end{center}
\caption{Calculated atom and spin-resolved density of states (DOS)
for the Mn$_2$VAl and Mn$_2$VSi compounds when they are
half-metallic.} \label{fig3}
\end{figure}

But even if these compounds are not half-metallic in their
equilibrium lattice constants, small expansion of the lattice
parameter restores the half-metallicity. In figure \ref{fig3} we
have drawn the spin-resolved density of states (DOS) projected on
each atom for Mn$_2$VAl and Mn$_2$VSi compounds. The rest of the
compounds present DOS's similar to their isoelectronic
counterparts. If we follow the nomeclature introduced in reference
\cite{GalanakisFull} to denote the spin-up and  spin-down bands in
Heusler alloys, for the Mn$_2$VZ compounds the situation is
reversed to all other full-Heusler alloys. Now the minority
spin-band is the spin-up one and the majority is the spin-down.
Thus the  gap is situated in the spin-up band and not the
spin-down one \cite{GalanakisFull}. This explains also why in
table \ref{table1} the total spin moment is negative. When we pass
from Mn$_2$VAl to Mn$_2$VSi which has one electron more we
populate one more spin-down state and the total spin moment
changes from -2 to -1 $\mu_B$. The charge of this electron is
attributed 1/3 to V and 2/3 to Mn. We also remark in figure
\ref{fig3} that the gap is larger for V and Al or Si atoms with
respect to the Mn ones. The states responsible for this differenc
are the $t_{1u}$ states of Mn just below the Fermi level (see
discussion in reference \cite{GalanakisFull}).

\begin{table}
\caption{Half metallic lattice parameters  and spin magnetic
moments (in  $\mu_B$) of Mn$_2$VZ (Z=Al, Si, Ge, Sn ). }
\begin{indented}
 \item[]
 \begin{tabular}{lccccc}
  \hline \hline
   Compound& $a(\AA)$ &  $m_\mathrm{Mn}$ & $m_\mathrm{V}$ &
$m_\mathrm{Z}$ & $m_\mathrm{Cell}$  \\
  \hline
Mn$_2$VAl   &5.987& -1.510  & 0.967 & 0.053    & -2.00 \\
            &6.117& -1.655  & 1.233 & 0.076    & -2.00 \\
 \hline
Mn$_2$VSi   &6.06 & -0.863  & 0.675 & 0.052    & -1.00 \\
            &6.29 & -1.092  & 1.105 & 0.078    & -1.00 \\
 \hline
Mn$_2$VGe   &6.18 & -0.976  & 0.905 & 0.048    & -1.00 \\
            &6.27 & -1.082  & 1.111 & 0.054    & -1.00 \\
 \hline
Mn$_2$VSn   &6.25 & -0.980  & 0.922 & 0.038    & -1.00 \\
            &6.31 & -1.050  & 1.059 & 0.041    & -1.00 \\
  \hline \hline
\end{tabular}
\end{indented}
 \label{table2}
\end{table}

In table~\ref{table2} we have gathered the half-metallic
parameters and the spin moments for four of the compounds under
study. We provide for each compound the smallest and largest
lattice constant for which half-metallicity is present. Of course
the heavier the system the largest the lattice parameters. We also
note that the range of paramters for which half-metallicity
appears is doubled in the case of Mn$_2$VAl(Si) with respect to
Mn$_2$VGe and even tripled with respect to Mn$_2$VSn. The lighter
the elements, the smaller the Coulomb respulsion and this leads to
larger gaps covering a larger area of lattice constants. Expansion
of the lattice leads in all cases to a small rearrangement of the
charge and the absolute values of the spin moments of all  atoms
increase. Especially in the case where the $sp$ belongs to the IVB
column the spin moment of V is comparable to the spin moment of
each Mn atom. Conlcuding we should note that half-metallicity is
feasible in these compounds depending on the lattice parameter
adopted and thus the choice of the proper substrate in experiments
is of primordial importance to get the highest possible
spin-polarization in experiments.

\section{Effect  of  disorder  on the half--metallicity: Mn$_2$V$_{1-x}$Al$_{1+x}$
and Mn$_2$V$_{1-x}$Si$_{1+x}$ alloys}
\label{sec5}

The effect of mixing the $sp$ atom has been already extensively
studied for the full-Heusler alloys and especially in reference
\cite{GalanakisQuat} it was shown that mixing the Al and Ge atoms
does not destroy the gap. On the contrary the
Mn$_2$V[Al$_x$Ge$_{1-x}$] remains half-metallic for any
concentration of Al and Ge atoms. This implies that it also shows
the Slater-Pauling behaviour and the total spin moment scales
following the rule $Z_t-24$. In our study we will seek the
influence of mixing now the lower-valent transition metal atom (V)
and the $sp$ atom. We took as prototypes Mn$_2$VAl and Mn$_2$VSi
at such a lattice constant that the Fermi level lies in the middle
of the gap, and then studied the families
Mn$_2$V$_{1-x}$Al$_{1+x}$ and Mn$_2$V$_{1-x}$Si$_{1+x}$ for
$x\epsilon [-0.2,0.2]$ with a step of 0.05.

\begin{figure}
\begin{center}
\includegraphics[scale=0.5]{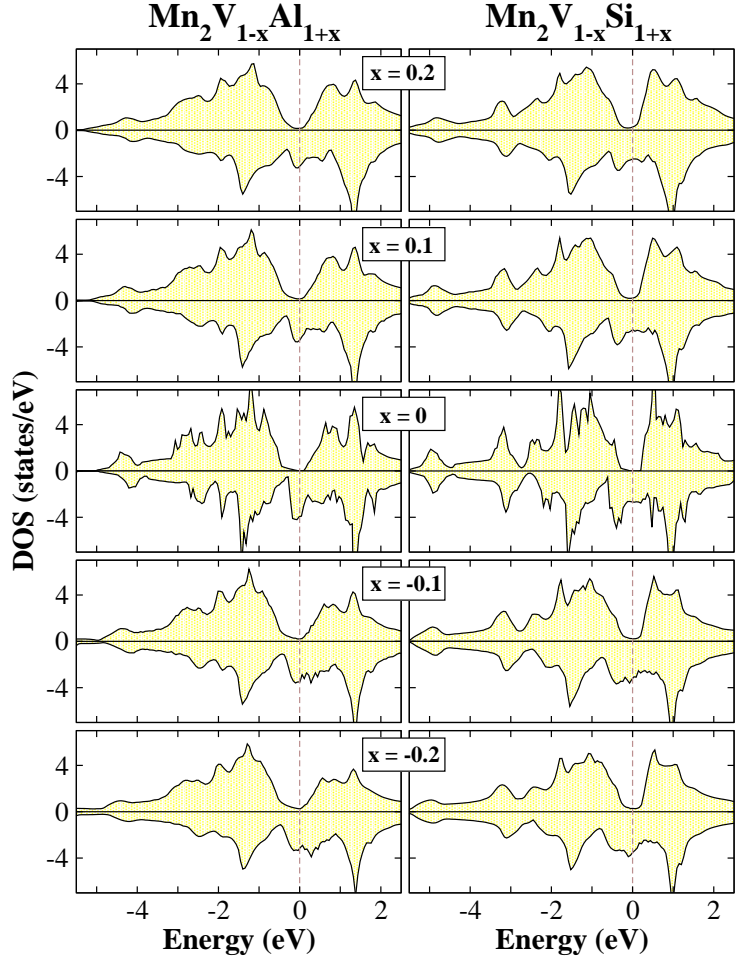}
\end{center}
\caption{Spin-resolved DOS for the Mn$_2$V$_{1-x}$Al$_{1+x}$ (left
pannel) and Mn$_2$V$_{1-x}$Si$_{1+x}$ (right pannel) compounds.}
\label{fig4}
\end{figure}

\begin{figure}
\begin{center}
\includegraphics[scale=0.5]{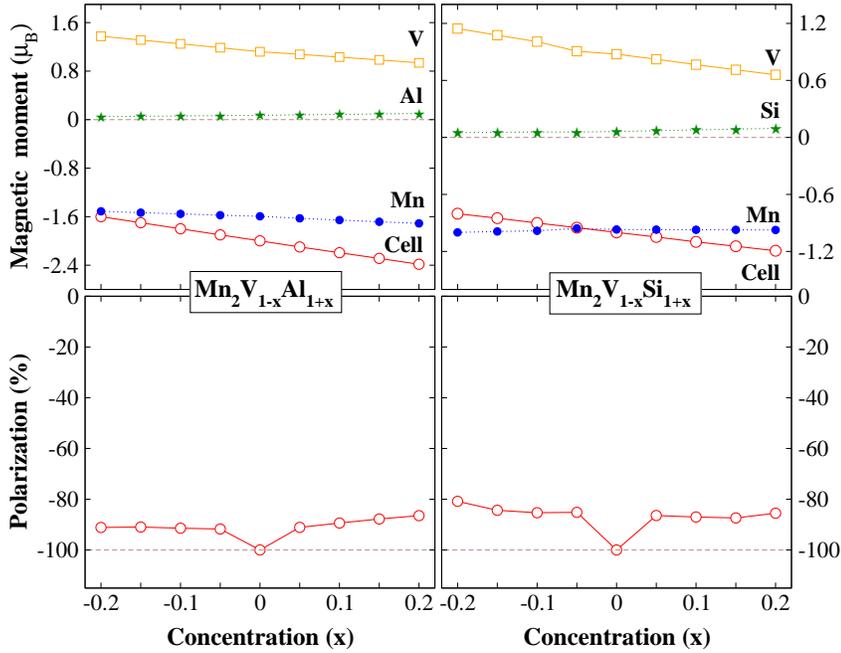}
\end{center}
\caption{Variation of the spin magnetic moments and the
spin-polarisation at the Fermi level with the concetration $x$ for
the Mn$_2$V$_{1-x}$Al$_{1+x}$ and Mn$_2$V$_{1-x}$Si$_{1+x}$
compounds.} \label{fig5}
\end{figure}

In figure~\ref{fig4} we have gathered the spin-resolved DOS for
all cases under study. The left pannel contains the resutls for
the Al-based compounds. We note that for $x=0$ we have a
half-metallic compound as already stated. When we have a surplus
of Al atoms ($x$=0.1 and 0.2) then the gap shrinks to almost zero
but the Fermi level continues to fall within it. An excess of V
atoms ($x$=-0.1 and -0.2) has a similar effect. The physics is
even clearer when we look at figure~\ref{fig5} where we represent
the variation of the spin moments and spin polarisation with the
concentration. When we are away from $x$=0 the system keeps a high
degree of spin-polarisation of about -90\%, which means that
almost all electrons ($\sim$95\%) are of spin-down character. The
spin-magnetic moment of the Mn atom is not influenced by the
disorder and remains almost constant for all values of $x$. The
spin moment of Al is almost negligible and has no real effect on
the total spin moment. It is the spin moment of V which changes
but as we can see from the total spin moment in the cell, it
follows the Slater-Pauling behaviour. For $x$=-0.2 we have the
Mn$_2$V$_{1.2}$Al$_{0.8}$ compound which has in average 22.4
electrons per unit cell and thus the total spin moment should be
-1.6 $\mu_B$ for the half-metallic case. For $x$=0.2, the compound
Mn$_2$V$_{0.8}$Al$_{1.2}$ has in average 21.6 valence electrons
and thus half-metallicity corresponds to -2.4 $\mu_B$. These
values coincide with the calculated one presented in figure
\ref{fig5}.

In the case of the Mn$_2$V$_{1-x}$Si$_{1+x}$ compounds the effect
of disorder is much larger and as we can see from
figure~\ref{fig5} away from the half-metallicity the spin
polarization is as low as -80\% and thus $\sim$90\% of electrons
at the Fermi level are of spin-down character. But the system
remains almost half-metallic as we can conclude by studying the
variation of the total spin moment. The values $x$=-0.2 and
$x$=0.2 correspond to 23.2 and 22.8 total valence electrons in the
cell in average, respectively. The calculated total spin moments
are about 0.8 and 1.2 $\mu_B$ as expected from the Slater-Pauling
curve for the half-metallic systems. Thus although disorder leads
to a decrease of the spin-polarization, a small amount of disorder
keeps an almost half-metallic character in the system.

The case of Mn$_2$V$_{1-x}$Si$_{1+x}$ compounds is a very
intesting one. Decreasing slightly the concentration in Si keeps a
really high spin-polarization and in the mean time the total
spin-moment in the cell decreases reaching a value of less than 1
$\mu_B$ while the individual spin moments of V and Mn atoms stay
really high. To this respect the case of Mn$_2$V$_{1.2}$Si$_{0.8}$
is very near to what we can call an ideal half-metallic
compensated ferrimagnet. Moreover even in the ideal case,
half-metallicity would be lost in the case of surfaces or
interfaces \cite{GalanakisSurfInter} making such a system even
more attractive and a possible candidate for spintronic devices.

\section{Conclusions}\label{sec6}

We have studied the possibility of appearance of half-metallicity
in the case of the full-Heusler compounds Mn$_2$VZ where Z is an
$sp$ atom belonging to the IIIB or IVB column of the periodic
table. These compound show ferrimagnetism with the V and Z spin
moments being antiparallel to the Mn ones. Firstly we performed
total-energy calculations for both the non-magnetic and
ferrimagnetic configurations to find the stable magnetic
configuration and the equilibrium lattice constant. We found that
when Z is Al, Ga, In and Sn the ferrimagnetic is stable with
respect to the non-magnetic state, while for the Si and Ge
compounds at the equilibrium lattice constant we could converge
only the non-magnetic solution. A small expansion of the lattice
leads to the emergence of ferrimagnetism also in these compounds.

Although all compounds are not half-metallic at their equilibrium
lattice constant, small expansion of the lattice pushes the Fermi
level within the gap which is now situated in the spin-up band
contrary to all other full-Heusler alloys. The total spin moment
is -2 $\mu_B$ for the Mn$_2$VAl(Ga or In) compounds which have 22
valence electrons per unit cell and -1 $\mu_B$ for the
Mn$_2$VSi(Ge or Sn) compounds with 23 valence electrons. Thus
theses compounds follow the Slater-Pauling behaviour and the "rule
of 24" \cite{GalanakisFull}. The lighter the element and the
smaller the number of valence electrons, the wider is the gap and
the more stable is the half-metallicity with respect to the
variation of the lattice constant.

Finally we examined in the case of the half-metallic Mn$_2$VAl and
Mn$_2$VSi compounds the effect of intermixing V with the Al or Si
atoms. We found that a small degree of disorder decreases the
spin-polarization at the Fermi level from its ideal -100\% value
but since this effect is only local, the resulting alloys still
show an almost half-metallic behaviour.

A small degree of substitution of V atom for Si ones in Mn$_2$VSi
keeps a high degree of spin-polarization and as shown by the total
spin moment in the cell the compound stays almost half-metallic.
In the mean time the total spin moment considerably decreases and
the alloy is very near to what we can call an ideal half-metallic
compensated ferrimagnet which has several advantages for realistic
spintronic applications. Our work urges experimentalists to grow
new half-metallic alloys on suitable substrates to get new
candidates for spintronic applications.

\ack E\c S acknowledges the financial support of Bundesministerium
f\"ur Bildung und Forschung.


\section*{References}

\begin{thebibliography}{100}


\bibitem{ohno}
Ohno H 1998 Science \textbf{281} 951

\bibitem{Zutic}
\v{Z}uti\'c I,  Fabian J and  Das Sarma S 2004 \RMP \textbf{76}
323

\bibitem{deBoeck}
 de Boeck J,  van Roy W,  Das J,  Motsnyi V,  Liu Z,  Lagae L,
Boeve H, Dessein K and  Borghs G 2002 \textit{Semicond. Sci.
Tech.} \textbf{17} 342


\bibitem{deGroot}
de Groot R A, Mueller F M, van Engen P G and Buschow K H J 1983
\PRL \textbf{50} 2024

\bibitem{Ishida}
Ishida S,  Akazawa S, Kubo Y  and Ishida J 1982 \textit{J. Phys.
F: Met. Phys.} \textbf{12} 1111; Ishida S, Fujii S, Kashiwagi S
and  Asano S 1995 \textit{J. Phys. Soc. Jpn.} \textbf{64} 2152


\bibitem{Soulen}
Soulen Jr R J, Byers J M, Osofsky M S, Nadgorny B, Ambrose T,
Cheng S F, Broussard P R, Tanaka C T, Nowak J, Moodera J S, Barry
Q  and  Coey  J M D 1998 \textit{Science} \textbf{282} 85


\bibitem{Freeman}
Stroppa A,  Picozzi S,  Continenza A and  Freeman A J 2003 \PR  B
\textbf{68} 155203

\bibitem{Akai}
Akai H 1998 \PRL \textbf{81} 3002

\bibitem{Akinaga}
Akinaga H, Manago T and  Shirai M 2000 \textit{Jpn. J. Appl.
Phys.} \textbf{39} L1118

\bibitem{Webster}
Webster P J and Ziebeck K R A, in {\em Alloys and Compounds of
d-Elements with Main Group Elements. Part 2.}, edited by H.R.J.
Wijn, Landolt-B\"ornstein, New Series, Group III, Vol. 19,Pt.c
(Springer-Verlag, Berlin), pp. 75-184

\bibitem{lattice_match_1}
Xie J Q,  Dong J W, Lu J, Palmstrm C J  and  McKernan S 2001
\textit{Appl. Phys. Lett.} \textbf{79} 1003

\bibitem{lattice_match_2}
Kurfiss M and Anton R 2003 \textit{J. Alloy. Comp.}  \textbf{361}
36

\bibitem{itoh}
Itoh H,  Nakamichi T,  Yamaguchi Y and Kazama N 1983
\textit{Trans. Jpn. Inst. Met.} \textbf{24} 265

\bibitem{Yoshida}
Yoshida Y, Kawakami M,  Nakamichi T 1981 \JPSJ \textbf{50} 2203

\bibitem{jiang}
Jiang C,  Venkatesan M and  Coey J M D 2001 \SSC \textbf{118} 513

\bibitem{Ishida2}
Ishida S,  Asano S and  Ishida J 1984 \JPSJ \textbf{53} 2718


\bibitem{ruben}
Weht R and  Pickett W E 1999 \PR \textbf{60} 13006

\bibitem{ersoy}
\c Sa\c s\i o\~glu E,  Sandratskii L M and  Bruno P 2005 \JPCM
\textbf{17} 995

\bibitem{Leuken}
van Leuken H and de Groot R A 1995 \PRL \textbf{74}  1171

\bibitem{koepernik}
Koepernik K and Eschrig H 1999 \PR \textbf{59} 3174; Koepernik K,
Velicky B, Hayn R and  Eschrig H 1998 \PR B \textbf{58} 6944

\bibitem{perdew}
Perdew J P and  Wang Y 1992  \PR  B  \textbf{45}  13244

\bibitem{Harmon}
Rhee J Y and Harmon B N 2004 \PR B  \textbf{70} 094411

\bibitem{ersoy2}
\c Sa\c s\i o\~glu E,  Sandratskii L M and  Bruno P 2004 \PR B
\textbf{70} 024427

\bibitem{GalanakisFull}
Galanakis I, Dederichs P H and
  Papanikolaou N 2002 \PR B \textbf{66} 174429

\bibitem{GalanakisQuat}
Galanakis I 2004 \JPCM \textbf{16} 3089

\bibitem{GalanakisSurfInter}
Galanakis I, Le\v{z}ai\'c M, Bihlmayer G and Bl\"ugel S 2005 \PR B
{\bf 71}, 214431; Le\v{z}ai\'c M, Galanakis I, Bihlmayer G, and
Bl\"ugel S 2005 \JPCM {\bf 17} 3121; Galanakis I 2002 \JPCM
\textbf{14} 6329

\end{thebibliography}
\end{document}